\documentstyle[preprint,prl,aps,epsf,epsfig]{revtex}  
\begin{document}  
\draft  
\tighten  
  
\title{Study of thermometers for measuring  
a microcanonical phase transition in nuclear fragmentation}  
  
\author{A. Le F\`evre$^1$, O. Schapiro$^2$ and A. Chbihi$^1$}  
  
\address{  
$^1$ GANIL, BP 5027, F-14021 Caen-Cedex, France\\  
$^2$ Hahn-Meitner-Institut  
Berlin, Glienickerstr. 100, D-14109 Berlin, Germany\\  
}  
  
\date{\today}  
  
\maketitle  
  
\begin{abstract}  
The aim of this work is to study how the thermodynamic temperature  
is related to the known thermometers for nuclei especially in view of studying the  
microcanonical phase transition. We find within the MMMC-model  
that the "S-shape" of the   
caloric equation of state $e^*(T)$ which is the signal of a phase   
transition in a system with conserved energy, can be seen in the   
experimentally accessible slope temperatures   
$T_{slope}$ for different particle types and also in the isotopic temperatures  
$T_{He-Li}$. The isotopic temperatures  
$T_{H-He}$ are weaker correlated to the shape of the thermodynamic temperature   
and therefore are less favorable to study the signal of a microcanonical
phase transition. We also show that the signal is very sensitive to variations in
mass of the source.
 
\end{abstract}  
  
  
In this work we are interested in testing the different experimentally   
accessible thermometers for nuclei in order to understand which quantity   
is best related to their thermodynamic temperature. It is our purpose to show  
that it should be in principle possible to measure the specific signal of a   
microcanonical phase transition in an accurate experiment.  
  
The concept of phase transitions is usually discussed in connection with   
macroscopic systems. In such systems phase transitions are recognized  
by divergences in quantities like specific heat $c(e^*)= de^*/dT_{thd}$,   
where $e^*$ is the specific excitation energy and $  
T_{thd}$ the thermodynamic temperature.  
If we are interested by similar phenomena in finite systems, the divergences  
are unsuitable to recognize and to classify phase transitions,   
since no divergences can occur. In a finite system the conservation laws   
become significant for the appearance and the shape of a phase transition.   
If a finite system is in thermal contact with a heat bath of constant   
temperature $T_{thd}$, i.e. is a canonical  
system, then the signal of a first order phase transition will become smeared   
showing up as a bump (instead of a divergence) in heat capacity or,   
equivalent, as a smooth anomaly in the caloric curve $e^*(T_{thd})$.  
For a finite and {\em isolated}, i.e. microcanonical system the signal 
changes  
qualitatively. Here the total energy $E$ of the system is a strictly  
conserved quantity. The first order phase transition becomes signaled by  
an "S-shape" in the caloric equation of state $T_{thd}(E)$ 
\cite{gross96,hueller94a}, as shown in   
the following figures in this publication.   
  
Since excited nuclei are an example of finite as well as isolated systems,  
they are especially suitable to study the signal of a microcanonical   
phase transition. This signal with an "S-shape" was frequently obtained in   
calculations, though it is still a subject of controversial discussion.  
Therefore it is of fundamental interest to test   
the theoretical findings by an experiment.   
   
Another important point is the fact that the thermodynamic temperature which  
can be simply measured in macroscopic physics is not accessible directly  
in a finite system. In a macroscopic system the measurement is simple through  
the fact that the size of the thermometer which gets into thermal contact  
with the system of interest is negligible compared to the size of the   
system.  
For a finite system this is obviously not satisfied. Especially for  
a microcanonical system it is not possible to bring a thermometer  
into thermal contact with a system without violating the strict energy   
conservation. Thus to obtain information   
on the thermodynamic temperature one needs to find an experimental observable   
which is not a temperature, but keeps information on the behavior of $T_{thd}$.  
  
Several suggestions for such observables were made.  
The candidates for "nuclear thermometers" are the slope temperatures from  
Maxwellian fits of energy distributions and the temperatures deduced   
from the isotopic ratios. In this work we are testing the quality of  
these "nuclear thermometers" as concerning their ability to reproduce   
the shape of the microcanonical caloric equation of state ($CES$)   
$T_{thd}(E)$. Another approach can be found in \cite{gulminelli97}.
  
We concentrate on the microcanonical phase transition from evaporation to   
asymmetric fission which was predicted in the fragmentation of hot nuclei   
within the Berlin statistical fragmentation model MMMC \cite{gross90}. 
Similar signals were also predicted by 
other statistical models for nuclear fragmentation \cite{bondorf95,sneppen}
and also for atomic clusters \cite{gross97,bixon89,mydiss}.  
An experimental observation in ref. \cite{chbihi98} seems also to support the  
existence of this phase transition. 

To study the "nuclear thermometers" we first obtain the  signal   
of a phase transition in the $T_{thd}(E)$ curve within the MMMC-model.   
Then we calculate the signals of the two thermometers in question, the 
caloric curves of $T_{slope}(E)$ and $T_{isotopic}(E)$.  
  
Let us first briefly discuss the basics of the model.  
The strictly microcanonical MMMC-model \cite{gross90} assumes a hot 
compound nucleus to be formed in a  
nuclear collision. After fragmentation of this compound system   
the fragments remain coupled and exchange nuclei as long as they are in close  
contact. Consequently, the system is assumed to equilibrate statistically  
shortly after the break-up. The volume which is accessed by the equilibrated  
fragment configuration is called the  
freeze-out volume. This means in terms of thermodynamics that   
the collection of all possible fragment configurations represents the   
maximum accessible phase space $\Omega(E)$ for a given   
freeze-out volume. $\Omega(E)$ is restricted  
by the valid conservation laws, which are the conservation of mass, charge, linear 
and angular momentum and of the total energy of the system, and also by   
the geometrical constraints.   
In the simulation the most important geometrical constraint is the size of the   
freeze-out volume, which is taken to be spherical.  
The radius $r_f$ of this volume, which is the only simulation parameter of the  
MMMC-model, is for the energy region of 1 to 4 MeV per nucleon   
at about $r_f=2.2A^{1/3}$fm,   
which corresponds to approximately 6 times the normal nuclear volume.   
When the fragments (which can be in excited states)  
leave this volume they may de-excite as they trace out  
Coulomb-trajectories. 
  
The output of the MMMC calculation is a collection of freeze-out configurations  
which are supposed to be representative for the entire phase space $\Omega(E)$.  
The thermodynamic temperature $T_{thd}$ of these configurations is calculated   
by 
\begin{equation}  
\frac{1}{T_{thd}}=\frac{\partial S}{\partial E}=\frac{\partial s}{\partial E^*}~~,  
\end{equation}   
where $E$ is the total energy of the system, $S=k_B \ln \Omega(E)$ is the entropy, 
$k_B$ the Boltzmann constant and $s=S/A$ and $E^*=E/A$ with $A$ the mass of the 
decaying nucleus.  
The caloric curve $T_{thd}(E^*)$ is plotted in all figures of this paper as a   
solid line with circles.  
  
We test the "nuclear thermometers" in two steps. First we plot the caloric   
curves $T_{slope}(E^*)$ and $T_{isotopic}(E^*)$ obtained   
from the MMMC-events after performing the Coulomb trajectories.  
Next we subject the calculated events to the software filter of the  
INDRA setup \cite{INDRA_filter}  and plot the filtered caloric curves. 

For the INDRA-filter we assume the source velocity as 8.1 cm/ns,
which is a typical quasi-projectile velocity measured with INDRA in 
mid-peripheral  Xe~+~Sn collisions at 50 A.MeV bombarding energy \cite{lefevre97}.
After the filtering we select the complete events for which the detected total 
charge and the total momentum is greater than 80\% of the initial charge and 
of the initial momentum of the source, respectively. These events are used for 
constructing the filtered caloric curves.
  
We start with the slope temperatures $T_{slope}$ for protons, deuterons,   
tritons, $^3He$ and alpha particles.  
The calculated kinetic energy spectra were fitted with the  
surface-evaporating Maxwellian source formula \cite{Awes} for every particle type:  
\begin{equation}  
\frac{d\sigma}{dE_{kin}}\propto  
\:  
\frac{(E_{kin}-B)}{T_{slope}^2}   
e^{(-\frac{E_{kin}-B}{T_{slope}})},  
\end{equation}  
where $E_{kin}$ is the center of mass kinetic energy of the particles, $B$  the  
Coulomb barrier and $T_{slope}$, which is the slope of the raw spectra, which   
is the desired slope temperature. We calculate the slope temperature for protons, 
deuterons, tritons $^3$He, alpha and also light IMFs. 

%Let us say several words on the philosophy of this widely applied formula:  
  
\begin{figure}  
\begin{center}
\leavevmode
\epsfxsize=17cm
\epsfbox{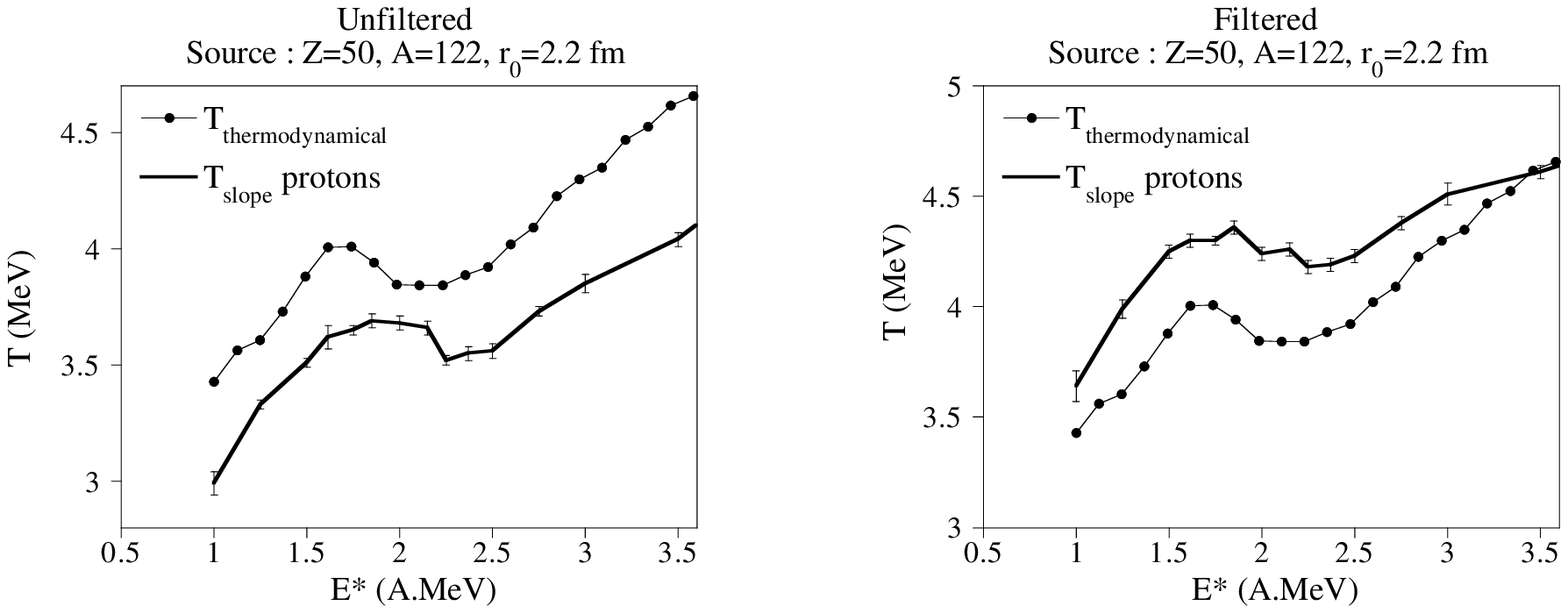}
\vspace{.7cm}
\caption{Caloric curves $T_{thd}(E^*)$ and $T_{slope}(E^*)$ for protons.  
Left panel: MMMC calculation, $T_{slope}(E^*)$ unfiltered.  
Right panel: MMMC calculation, $T_{slope}(E^*)$ filtered according to the   
INDRA setup.  
}  
\label{slopep}  
\end{center}  
\end{figure}

\begin{figure}  
\begin{center}
\leavevmode
\epsfxsize=17cm
\epsfbox{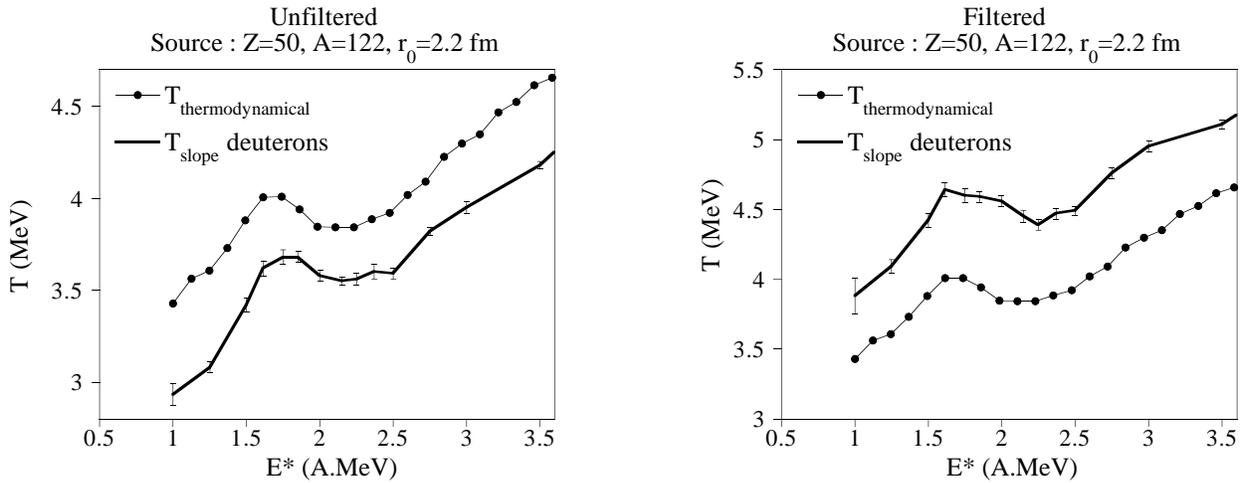}
\caption{  
Like figure~\protect\ref{slopep}\protect, but for deuterons.  
}  
\label{sloped}
\end{center}  
\end{figure}

\begin{figure}  
\begin{center}
\leavevmode
\epsfxsize=17cm
\epsfbox{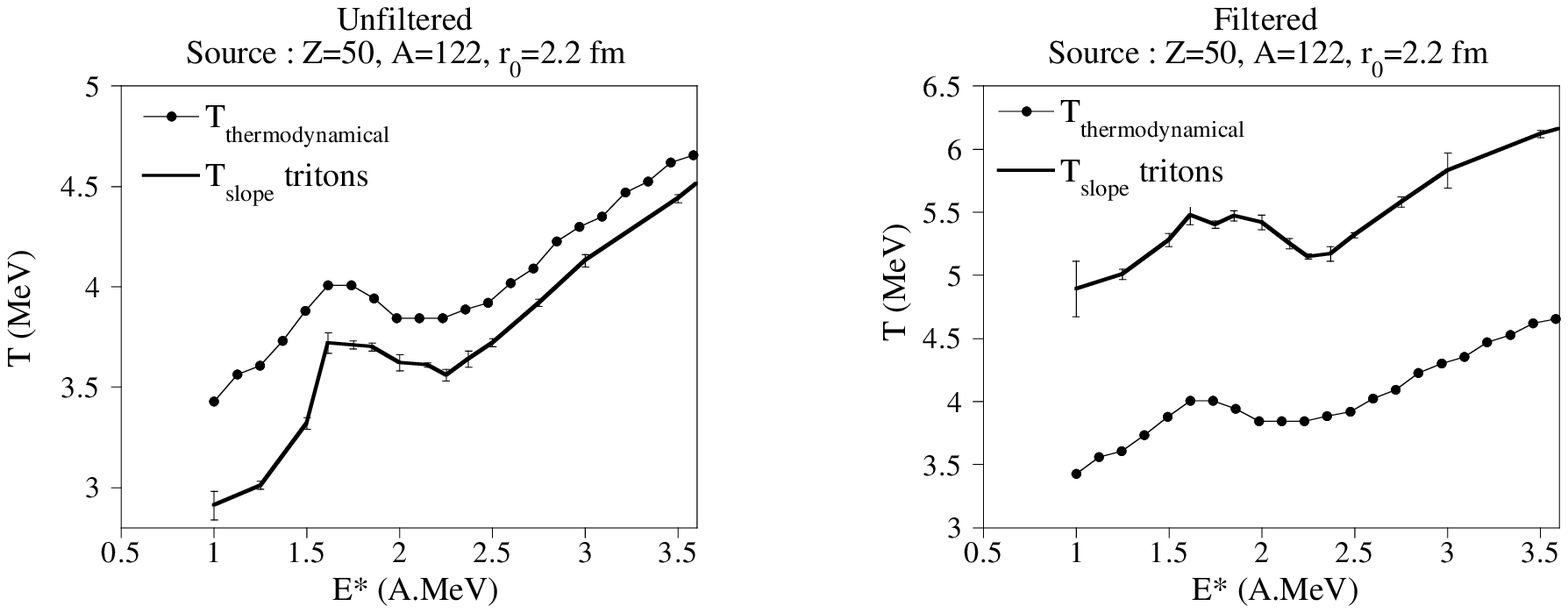}
\caption{  
Like figure~\protect\ref{slopep}\protect, but for tritons.  
}  
\label{slopet}  
\end{center}  
\end{figure}

\begin{figure}  
\begin{center}
\leavevmode
\epsfxsize=17cm
\epsfbox{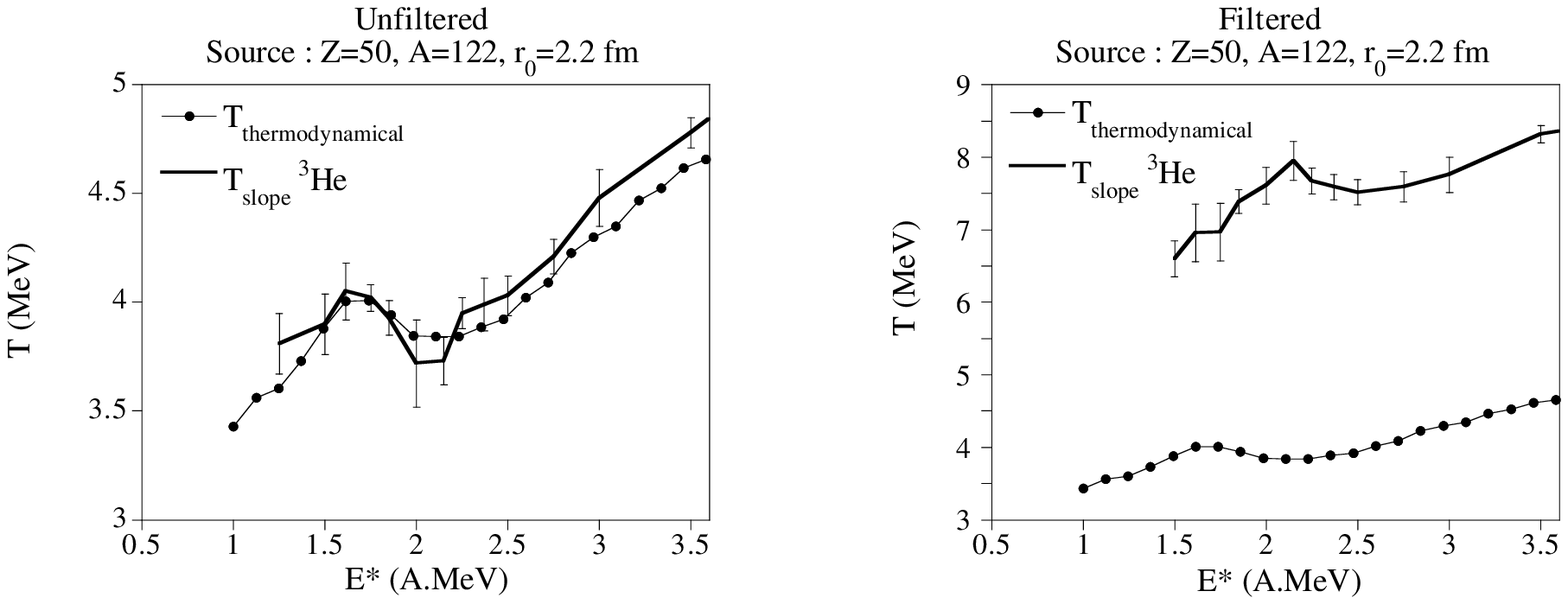}
\caption{  
Like figure~\protect\ref{slopep}\protect, but for $^3He$  
}  
\label{slopehe3}  
\end{center}  
\end{figure}

\begin{figure}  
\begin{center}
\leavevmode
\epsfxsize=17cm
\epsfbox{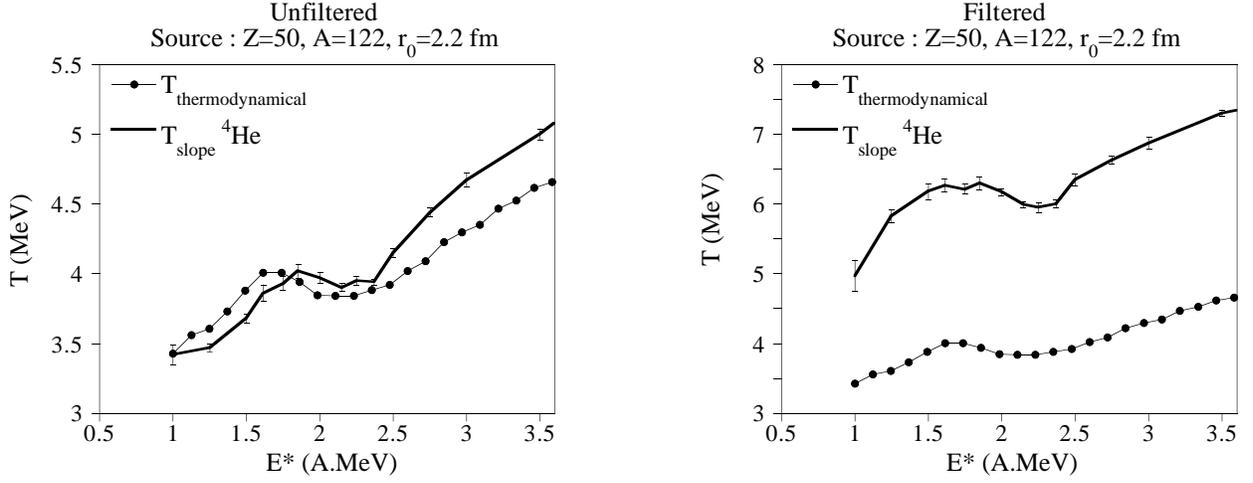}
\caption{  
Like figure~\protect\ref{slopep}\protect, but for alpha.  
}  
\label{slopea}  
\end{center}  
\end{figure}

Figures~\ref{slopep}, \ref{sloped}, \ref{slopet}, \ref{slopehe3} and   
\ref{slopea} show the   
comparison of $T_{thd}(E^*)$ and $T_{slope}(E^*)$ for unfiltered   
(left plot) and INDRA-filtered (right plot) events for $p$, $d$, $t$,   
$^3He$ and $\alpha$.    
Our most important finding in all these curves is   
that the slope temperatures resemble the general shape of $T_{thd}(E^*)$ before   
and after the filtering. 

The caloric curves for Li, Be and B (not shown here) which we have calculated only 
without filtering, repeat also the general shape of the phase transition despite of 
big error bars in $T_{slope}$.
  
The unfiltered $T_{slope}(E^*)$ for $p$, $d$, $t$  
systematically achieve values lower then the thermodynamic temperatures, while  
the values for $^3He$ and $\alpha$ are close to those of $T_{thd}$. We think   
that the last finding is just accidental. The   
unfiltered $T_{slope}$ are close for all particles with $Z=1$ and for those  
with $Z=2$, and the shift between them of $\approx 0.3$ MeV is due to the   
higher Coulomb repulsion. Here one can see in a very simple way that the   
result of the MMMC calculation cannot be described by a Maxwellian  
source with a {\sl unique} temperature $T_{slope}(E^*)$. 
Still we see that using the Maxwellian   
fit just as a recipe 
%without further physical interpretation 
we obtain a   
useful tool to extract a pseudo-temperature which is correlated with the   
thermodynamic temperature.   
    
Let us now proceed to the isotopic temperatures \cite{albergo85}.  
The basic assumption for the isotopic temperature formula is a thermal equilibrium
between free nucleons and composite fragments within a certain interaction volume V
at a temperature T. The formalism is of the grandcanonical ensemble and ignores the 
possible effects of mass and energy conservation.

The $T_{He-Li}(E^*)$ isotopic temperature is given by
\begin{equation}  
T_{He-Li}=16/ln(2.18\frac{Y_{^6Li}/Y_{^7Li}}{Y_{^3He}/Y_{^4He}})~~, 
\label{T_He_Li}  
\end{equation}  
and the $T_{H-He}(E^*)$ temperature by  
\begin{equation}  
T_{H-He}=14.3/ln(1.6\frac{Y_{^2H}/Y_{^3H}}{Y_{^3He}/Y_{^4He}})~~,  
\label{T_H_He}  
\end{equation}  
where $Y(E^*)$ is the particle yield.

\begin{figure}  
\begin{center}
\leavevmode
\epsfxsize=11cm
\epsfbox{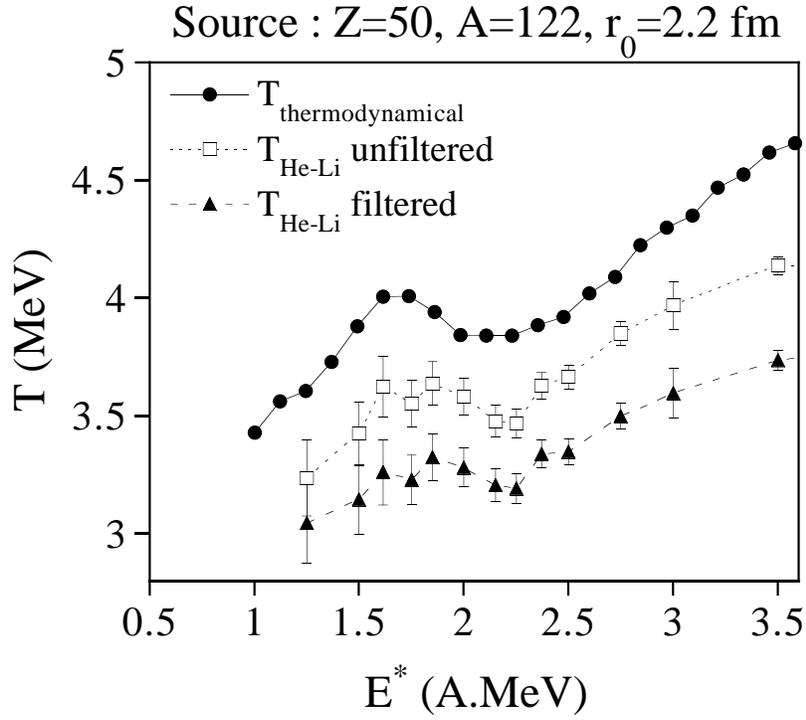}
\vspace{.5cm}
\caption{  
Caloric curves $T_{thd}(E^*)$ and $T_{He-Li}(E^*)$ from the MMMC calculation.  
$T_{He-Li}$ is shown unfiltered and filtered according to the  
INDRA setup.  
}  
\label{isoHeLi}  
\end{center}  
\end{figure}

Fig.~\ref{isoHeLi} shows a comparison of unfiltered and filtered $T_{He-Li}(E^*)$  
with the thermodynamic temperature $T_{thd}(E^*)$. Again we see that the signal  
of the phase transition survived the procedure.  
  
\begin{figure}  
\begin{center}
\leavevmode
\epsfxsize=11cm
\epsfbox{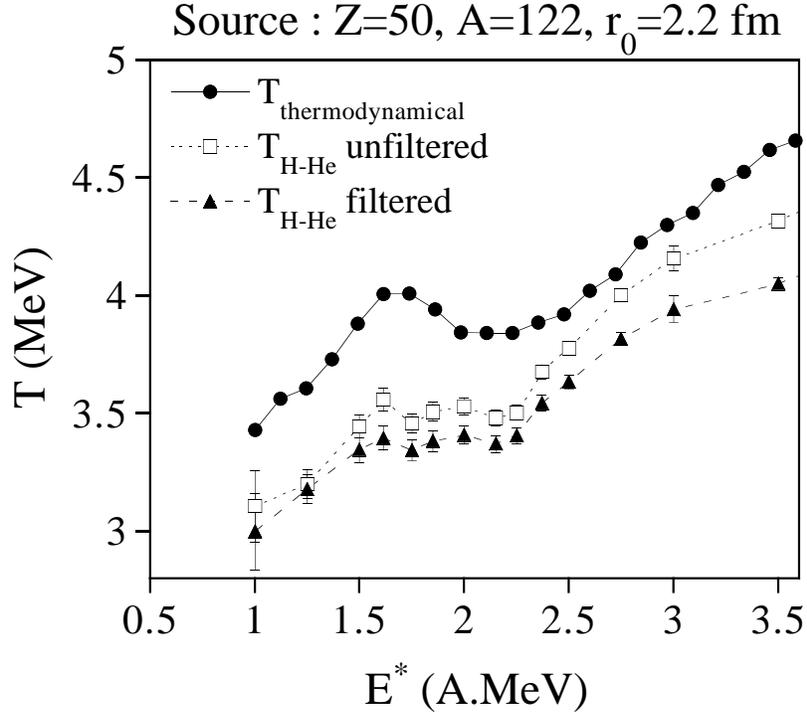}
\caption{Like figure~\protect\ref{isoHeLi}\protect, but for H-He isotopic   
temperature $T_{H-He}(E^*)$.}  
\label{isoHHe}  
\end{center}  
\end{figure}

Performing the same for the H-He isotopic ratios, figure~\ref{isoHHe}, we find   
that already the unfiltered $T_{H-He}(E^*)$ isotopic temperature is less 
sensitive to the "S-shape" in the $T_{thd}(E^*)$ curve, but it also shows some
structure at the phase transition. 
  
\begin{figure}  
\begin{center}
\leavevmode
\epsfxsize=15cm
\epsfbox{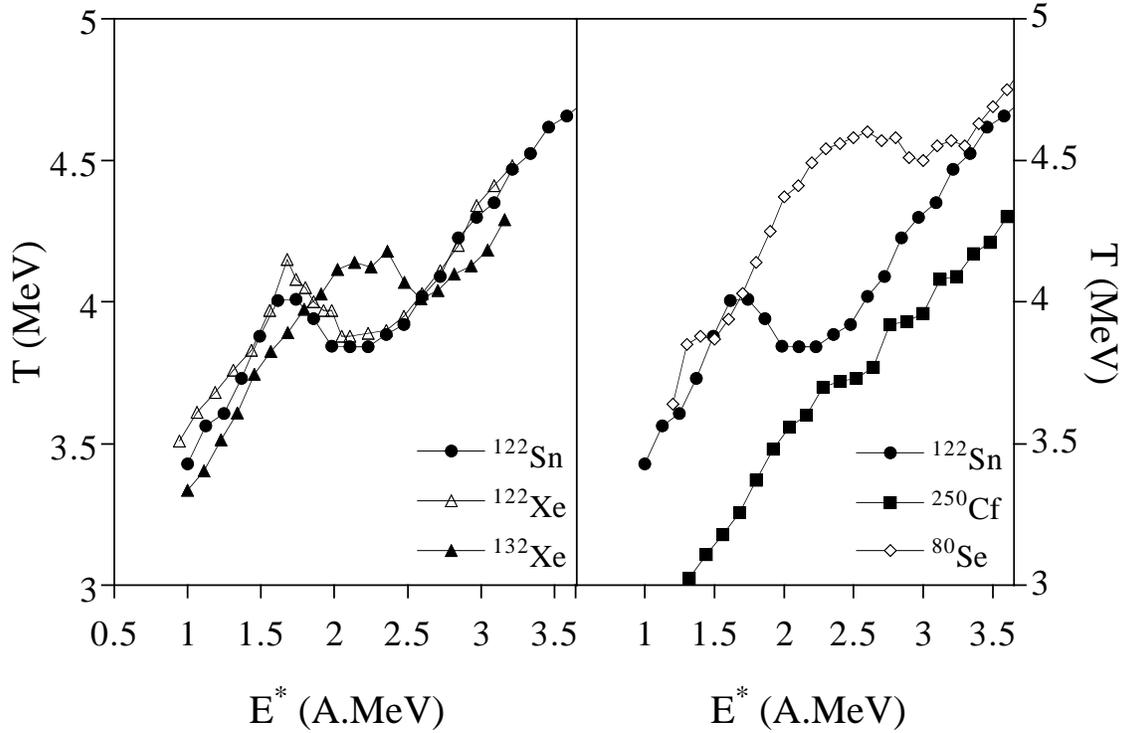}
\vspace{.5cm}
\caption{Signals of phase transitions in the thermodynamic  temperature
$T_{thd}(E^*)$ from MMMC calculations for different masses and charges of 
compound nuclei.}  
\label{Tthd}  
\end{center}  
\end{figure}

Finally we would like to address the question, how sensitive is the signal   
of the microcanonical phase transition to the mass and charge of the compound  
nucleus. Figure~\ref{Tthd} shows the $T_{thd}(E^*)$ curve for several   
masses and charges of the source. In the left panel we show that decreasing 
the charge of the source from Z=54 for $^{122}$Xe (dots)
to Z=50 for $^{122}$Sn (empty triangles) does not influence much the phase 
transition.
On the opposite, increasing the mass by 10 nuclei for $^{132}$Xe (full triangles)
shifts the transition 
signal to higher excitation energies by approximately 0.5 MeV per nucleon. 

In the right panel we show two additional curves for $^{80}$Se and $^{250}$Cf, 
thus strongly reducing and strongly decreasing the mass.
Here the evaporation-fission phase transitions for $^{80}$Se appears as a 
very weak signal at $E^* \approx 1.5$ A.MeV and $T_{thd} \approx 3.8$ MeV. The 
stronger
signal at $T_{thd} \approx 4.6$ MeV is due to a different fission-multifragmentation
phase transition. 
 
The fragmentation behavior of $^{250}$Cf shows no phase transition at all, even for 
higher or lower $E^*$. This is connected to the intrinsic instability against 
fission, so that 
even at very low excitation energies the nucleus fissions symmetrically instead of 
entering first the evaporation mode and later the asymmetric fission mode 
like the lighter nuclei do. This shows very plastically that there is no unique
liquid-gas phase transition in nuclear fragmentation, but many different transitions
depending on mass, and may be on other characteristics of nuclei.

The dependence of the phase transition on the mass is natural, 
since increasing the mass we automatically increase the phase 
space $\Omega(E^*)$. Thus if changing $A$ with excitation energy  
can also produce unusual shapes of the $T_{thd}(E^*)$ curve\footnote{We  
suppose together with \cite{natowitz95,ma97} and in opposite to 
\cite{lee97} that the curve shown in 
ref.~\cite{pochodzalla95} is just the   
effect of changing the mass of the source.} without undergoing any phase 
transition. This discussion   
serves mainly to make the point that to measure the microcanonical phase   
transition it is essential to keep especially the mass of the source at  
constant over the whole energy range.  
  
Further, it is important to obtain energy bining less than 0.2 MeV per nucleon to
detect the discussed transition. The investigation in ref. \cite{ma97} misses the 
phase transition by performing energy steps $\Delta E^* \approx 1$ A.MeV. 
The current INDRA setup can realize $\Delta E^* \approx 0.4$ A.MeV 
for the energy region of 1 to 4 A.MeV, which is at the limit of desired accuracy. 
If an additional experimental setup could measure the mass of the projectile 
fragments one could obtain 
much better $E^*$ reconstructions and $\Delta E^*$ below 0.2 A.MeV. 
  
Summarizing the above we studied how the shape of the caloric   
curve $T_{thd}(E^*)$ at a microcanonical phase transition is correlated    
to the shape of different nuclear thermometers $T_{slope}(E^*)$ and   
$T_{isotopic}(E^*)$. The signal of a microcanonical phase transition is an  
"S-shape" in the thermodynamic temperature. The slope temperatures for   
proton, deuteron, triton, $^3He$ and alpha and the isotopic temperature  
$T_{He-Li}(E^*)$ reproduce this "S-shape" at correct excitation energies  
for INDRA-filtered and unfiltered events.   
The absolute values of this curves vary and do not coincide with the   
thermodynamic temperature. This means that the decrease of temperature with  
rising excitation energy at a first order microcanonical phase transition   
can be measured experimentally. This is of fundamental importance   
for the systematic study of phase transitions in finite systems.

O.S. is grateful to GANIL for the friendly atmosphere during her stays there.   
This work was supported by IN2P3.

% now the references. delete or change fake bibitem. delete next three  
% lines and directly read in your .bbl file if you use bibtex.  

%\listoffigures  
  
%%\pagebreak  
  
%%{\bf Figure Captions}  
  
%%\normalsize  
  
\end{document}